# Robust Weyl Semimetal and Pressure-induced Superconductivity in Quasi-One-Dimensional $(NbSe_4)_2I$

Cuiying Pei[1#], Wujun Shi[2,3#], Yi Zhao[1], Lingling Gao[1], Jiacheng Gao[4,5], Yiwei Li[1,6], Haiyin Zhu[1], Qing Zhang[1], Na Yu[1], Changhua Li[1], Weizheng Cao[1], Sergey A. Medvedev[7], Claudia Felser[7], Binghai Yan[8], Zhongkai Liu[1], Yulin Chen[1,6], Zhijun Wang[4,5*], Yanpeng Qi[1*]

1. School of Physical Science and Technology, ShanghaiTech University, Shanghai 201210, China
2. Center for Transformative Science, ShanghaiTech University, Shanghai 201210, China
3. Shanghai high repetition rate XFEL and extreme light facility (SHINE), ShanghaiTech University, Shanghai 201210, China
4. Beijing National Laboratory for Condensed Matter Physics, and Institute of Physics, Chinese Academy of Sciences, Beijing 100190, China
5. University of Chinese Academy of Sciences, Beijing 100049, China
6. Department of Physics, Clarendon Laboratory, University of Oxford, Parks Road, Oxford OX1 3PU, UK
7. Max Planck Institute for Chemical Physics of Solids, 01187 Dresden, Germany
8. Department of Condensed Matter Physics, Weizmann Institute of Science, Rehovot 7610001, Israel

\# These authors contributed to this work equally.
\* Correspondence should be addressed to Y.Q. (qiyp@shanghaitech.edu.cn) or Z.W. (wzj@iphy.ac.cn)

**A search for the single material system that simultaneously exhibits topological phase and intrinsic superconductivity has been largely limited, although such a system is far more favorable especially for the quantum device applications. Except artificially engineered topological superconductivity in heterostructure systems, another alternative is to have superconductivity arising from the topological materials by pressure or other "clean" technology. Here, based on first-principles calculations, we first show that quasi-one-dimensional compound $(NbSe_4)_2I$ represents a rare example of a chiral Weyl semimetal in which the set of symmetry-related Weyl points (WPs) exhibit the same chiral charge at a certain energy. The net chiral charge (NCC) of the below Fermi level $E_F$ (or a certain energy) can be tuned by pressure. In addition, a partial disorder induced by**

**pressure accompanied with superconductivity emerges. Although amorphization of the iodine sub-lattice under high pressure, the one-dimensional NbSe$_4$ chains in (NbSe$_4$)$_2$I remain intact and provide a superconducting channel in one dimension. Our combined theoretical and experimental research provide critical insight into a new phase of the one-dimensional system, in which distinctive phase transitions and correlated topological states emerge upon compression.**

The physics of one-dimensional or quasi-one-dimensional materials is the subject of very active research due to their very peculiar electronic properties, which are dramatically different from two- or three-dimensional systems[1-4]. The quasi-one-dimensional compounds can be regarded as being composed of chains, the internal cohesion of which is due to strong iono-covalent or metallic bonds. In contrast, interchain bonds are very weak and generally of the wan der Waals type. When the motion of electrons is confined within one dimension, the electrons cannot move without pushing all the others, which leads to a collective motion and thus, spin-charge separation. In such a case, a Tomonaga–Luttinger liquid (TLL) is formed, where single-particle excitations are replaced by highly correlated collective excitations[5].

Iodized transition-metal tetrachalcogenides (MSe$_4$)$_2$I (M = Ta, Nb) belong to quasi-one-dimensional compounds[6-8]. Since its first synthesized in 1982[8], (TaSe$_4$)$_2$I have been widely studied. Very recently, our transport measurements clearly indicated an exotic axionic charge-density wave (CDW) state in quasi-one-dimensional compound (TaSe$_4$)$_2$I[9]. Furthermore, we employ both first-principles calculations and experimental probes to explore the electric structure and approve that (TaSe$_4$)$_2$I hosts topological chiral fermions at $E_F$, and the first reported link between the bulk WPs and the CDW wavevector[10]. Compared with (TaSe$_4$)$_2$I[11-15], the sister compound (NbSe$_4$)$_2$I is far from explored both theoretically and experimentally[16-17].

Pressure is a unique technique to tune the interchain hopping and then many interesting physical phenomena emerge, such as superconductivity[18-30]. In fact, pressure-induced superconductivity has been experimentally observed in the compound

$(TaSe_4)_2I^{31-33}$. A question arises naturally: is $(NbSe_4)_2I$ also a Weyl semimetal at ambient conditions and is it possible to achieve superconductivity under high pressure while their topological states remain intact. Here, we systematically investigate the high-pressure behavior of the novel quasi-one-dimensional $(NbSe_4)_2I$, which is a rare example of a chiral Weyl semimetal. Through *ab initio* band structure calculations, we find that in the absence of spin orbit coupling (SOC), the symmetry-protected WPs[34] are pinned at the high-symmetry point —P. Upon considering SOC, these WPs still survive with double total chiral charge. In addition, the NCC of the WPs below $E_F$ can be tuned by external pressure, since its crystals have the chiral symmetry. Our synchrotron X-ray diffraction (XRD) and Raman spectroscopy results demonstrate that the $NbSe_4$ building blocks of the quasi-one-dimensional chains survive under high pressure, while the rest of the structure becomes amorphous. Interesting, upon the amorphization of the atomic chains, pressure-induce metallization and superconductivity emerged in our novel one-dimensional chain system.

$(NbSe_4)_2I$ crystals used for the study were structurally characterized using single-crystal XRD, Energy dispersive X-ray spectroscopy (EDS) and high-angle annular dark-field scanning transmission electron microscopy (HAADF-STEM). The detailed results are shown in supplementary materials (Table S1-3, Figure S1-3). $(NbSe_4)_2I$ is a body-centered tetragonal chiral structure (space group *I*422, No. 97), as shown in Figure 1a. The conventional cell is composed by two primitive cell. The Brillouin zone of the primitive cell and the (001) projection are displayed in Figure 1c. The 1D building blocks of $(NbSe_4)_2I$, along the [001] and [110] axes, are revealed by HAADF-STEM images (Figure 1b). It consists of $NbSe_4$ chains parallel to *c* axis and separated by iodine atoms. In an $NbSe_4$ infinite chain, each metal is sandwiched by two rectangular selenium units. The dihedral angle between adjacent rectangles is about 45°, so that the stacking unit is an $MSe_8$ rectangular antiprism.

Owing to its quasi-1D crystal structure, $(NbSe_4)_2I$ exhibits a strongly anisotropic electronic structure. The first-principles calculations were carried out to obtain the electronic band structures of $(NbSe_4)_2I$ along high-symmetry lines without and with SOC. The band structures are shown in Figure 2. We observe weak dispersion in the $k_z$

= $\pi/c$ plane along *PN*, and observe very strong dispersion along *ΓZ*, as $k_z$ is along the chain direction. We also observe 10 (24) pairs of WPs without (with) SOC (in Table 1). The (001) view of the WPs distributions at the ambient pressure without and with SOC is shown in Figure 1d. The solid octagon indicates the projected boundary of the first bulk BZ, and the differently shaped symbols denote the WPs with different Chern numbers.

Due to the crystalline and time-reversal symmetry, it is a symmetry-protected Weyl semimetal (SPWSM) in (NbSe$_4$)$_2$I without SOC. In the absence of SOC, the charge-2 ($C = -2$) W2 (see Table 1) is pinned by the combined antiunitary symmetry—$TC_{4z}$—at the *P* point.[34] As thus, with infinitesimally SOC, it has to be a Weyl semimetal as well. Our calculations confirm that it belongs to a Weyl semimetal phase even with SOC, which is attributed to the large separation between the different-chiral-charged WPs $[|W1(C = -2) - W2(C = +2)| = \sqrt{2}\pi/a]$. The symmetry-protected Weyl point can be captured by the effective $k \cdot p$ model in the vicinity of the *P* point in the supplementary material (Table S4). Due to the lack of improper symmetry, such as the mirror and inversion symmetry, (NbSe$_4$)$_2$I represents a rare example of a chiral Weyl semimetal in which all of the symmetry-related WPs exhibit the same Chern number.

In order to confirm the Weyl semimetal properties, we calculated the Fermi surface of (001) surface at ambient pressure, the results are displayed in Figure 3a and d. The bulk states form several islands which have the net Chern number. The Fermi arcs are the striking properties of Weyl semimetals, which is related to the Chern number (Figure 1d and Figure 3). The calculated number of Fermi arcs are consistent with the calculate Chern numbers. We also carried out Angle-resolved photoemission spectroscopy (ARPES) measurement using a lab-based laser source, as shown in Figure S4. Equal energy mapping of the Fermi surface shows the quasi-one-dimensional electronic structure of (NbSe$_4$)$_2$I. Band dispersion was measured at both 82 K and 300

K, as shown in Figure S4b and Figure S4c, respectively. Above $T_C \sim 210$ K for the CDW phase[17, 35], the topological surface Fermi arcs is difficult to resolve due to temperature broadening[10]. A polaron gap of about 0.4 eV was identified and persisted under room temperature, which is similar to the observation in $(TaSe_4)_2I$[10]. A signature of band splitting at 300 K was observed which could be a result of surface charge polarization due to the loss of surface iodine atoms[15].

One-dimensional compounds are unstable and usually sensitive to pressure. We investigated the pressure-induced evolution of the structure of $(NbSe_4)_2I$ using *in situ* high-pressure XRD (Figure 4a and S5a). Our results indicate that one-dimensional structure of $(NbSe_4)_2I$ is robust in the low-pressure range ($P \leq 16.2$ GPa) (Figure 4a). Pressure-dependent Raman spectroscopy studies (Figure 4b and S5b) further demonstrate the stability of the $(NbSe_4)_2I$ structure up to 16 GPa, as no additional peaks or splitting are observed in the resulting spectra.

Next, we study the evolution of electronic structure with increasing pressure. The calculated electronic band structure without and with SOC under 5.0 GPa and 9.7 GPa are displayed in Figure 2b-c, and e-f, respectively. Due to the atom distance decrease with pressure increasing, the interaction between atoms increases and the band structure becomes more dispersive. Comparing the dispersion without and with SOC, the striking difference is the opened band gap along the *P-N* direction.

As the same as under the ambient pressure, our calculations show that even under different pressures (5.0 GPa, 9.7 GPa), it's still a SPWSM as long as the crystal symmetry is unchanged. When SOC is excluded, the W2 WPs are pinned at the *P* point (see Table 1), while SOC opens the band gap at *P* point (see Figure 2). It is interesting that the WP number decreases to 8 pairs under 5.0 GPa, and 16 pairs under 9.7 GPa with pressure increasing (see more details for the WPs in Table 1). Without SOC, the W4 WPs move towards the W2 WPs when the pressure increasing (see Figure 2 and Figure 3). The NCC of the lower WPs can be tuned by pressure; namely, the NCC of an iso-energy surface can be tuned effectively. $(NbSe_4)_2I$ sample should therefore exhibit a strongly amplified response in quantized bulk topological chiral probes, such as the chiral magnetic effect (CME) and the quantized circular photogalvanic effect (CPGE).

The calculated Fermi surface under 5.0 GPa and 9.7 GPa without and with SOC are displayed in Figure 3 b-c, and e-f, respectively. The Fermi arcs connected the separated islands which host the WPs with different Chern number.

It should be noted that structural disorder becomes apparent at pressures exceeding 16.2 GPa, and a new broad peak appears at ≈ 14.8° in the diffraction patterns (Figure 4a). This indicates that the applied high pressures introduced amorphous phases into the material. Similar phenomenon was observed in $(TaSe_4)_2I$[33]. Interestingly, the sharp profile of the (110) peak in the diffraction patterns is preserved throughout the structural evolution, and the $NbSe_4$ chains are the main blocks defining the (110) plane. Meanwhile, the strongest Raman peak (located at around 279 cm$^{-1}$) corresponding to the intrachain Se-Se stretching vibrations can be observed up to the highest pressure among these experiments (Figure 4b). Thus, the persistent of XRD diffraction (110) peak and isolated Raman peak demonstrate the preservation of the $NbSe_4$-units as building blocks of local order and the amorphization of the iodine sub-lattice in the high-pressure phase of $(NbSe_4)_2I$. The high-pressure phase, combination of one-dimensional chain and disordered sub-lattice, provides a novel phase of solid-state materials.

Next, we performed transport measurements on $(NbSe_4)_2I$ single crystal at various pressures. Figure 5a shows the temperature dependence of the resistivity for pressure up to 60.8 GPa. In the low-pressure range, the $\rho(T)$ displays a semiconducting-like behavior. With the pressure increasing, the overall magnitude of resistivity is suppressed continuously. As pressure increases up to 26.2 GPa, a small drop of resistivity in $(NbSe_4)_2I$ is observed at the lowest temperature (experimental $T_{min}$ = 1.8 K, shown in Figure 5b). Zero resistivity is obtained for pressure at 36.4 GPa, indicating the emergence of superconductivity. Although we did not detect resistivity anomaly in our high-pressure transport measurements due to smaller CDW gap, often superconductivity emerges in such low-dimensional system when CDW transition is suppressed by applied pressure[32, 36]. The critical temperature, $T_c$, keeps increasing monotonically and is 5.2 K at our maximum pressure of 60.8 GPa.

The appearance of bulk superconductivity in $(NbSe_4)_2I$ is further supported by the evolution of the resistivity-temperature curve with an applied magnetic field. Figure 5c displays $\rho(T)$ curves in external magnetic fields for $(NbSe_4)_2I$ at $P = 60.8$ GPa. It is clear that $T_c$ of $(NbSe_4)_2I$ is suppressed progressively by magnetic fields, and a magnetic field of $\mu_0 H = 5$ T deletes all signs of superconductivity above 1.8 K. We also tried to use the Ginzburg–Landau formula to fit the data: $\mu_0 H_{c2}(T) = \mu_0 H_{c2}(0)(1-t^2)/(1+t^2)$, where $t$ is the reduced temperature $T/T_c$. A critical field of $\mu_0 H_{c2} = 5.07$ T results for $(NbSe_4)_2I$, which yields a Ginzburg–Landau coherence length $\xi_{GL}(0)$ of 8.1 nm.

The measurements on different samples of $(NbSe_4)_2I$ provide the consistent and reproducible results, confirming this intrinsic superconductivity under pressure (Figure S6). The pressure dependences of the resistivity in the normal state at $T = 300$ K and of the critical temperature of superconductivity for $(NbSe_4)_2I$ are summarized in Figure 5d. At ambient pressure, $(NbSe_4)_2I$ is a chiral Weyl semimetal in which all of the symmetry-related WPs nearest the Fermi energy exhibit the same Chern number. Application of pressure could dramatically modify not only the electronic properties but also the net chiral charge (NCC) of the lower WPs. Above 16.5 GPa, a partial amorphization accompanied with a superconducting transition emerges. The $T_c$ continuously rises up to 5 K at the pressure of 60 GPa and still does not exhibit the trend of saturation. Since the amorphization of the iodine under high pressure, we cannot carry out band structure calculations. It should be noted that the Weyl points will be kept even though the I atoms was removed due to the symmetry protected Weyl points. The calculated band structure under different pressures with the I atoms were removed is displayed in Figure S7. It is clear that the band cross along $\Gamma Z$ direction is still kept, while the Fermi level moves to higher energy. This means that the topological properties should come from the $NbSe_4$ chain, while the I atoms accept the electrons. Thus, quasi-one-dimensional $(NbSe_4)_2I$ provides a promising platform for topological superconductor research.

In summary, we highlight that novel evolution of topological properties upon compression in (NbSe$_4$)$_2$I and the formation and quantum condensation of Cooper pairs in the one-dimensional NbSe$_4$ chains and the amorphization of the iodine sub-lattice. Iodized transition-metal tetrachalcogenides (MSe$_4$)$_2$I (M = Ta, Nb) compounds with nontrivial topology of electronic states display new ground states upon compression and potential applications to the next-generation quantum electronics devices.

**ACKNOWLEDGMENT**

This work was supported by the National Natural Science Foundation of China (Grant No. U1932217, 11974246 and 12004252), the National Key R&D Program of China (Grant No. 2018YFA0704300), Natural Science Foundation of Shanghai (Grant No. 19ZR1477300) and the Science and Technology Commission of Shanghai Municipality (19JC1413900). The authors thank the support from C$\hbar$EM (02161943) and Analytical Instrumentation Center (SPST-AIC10112914), SPST, ShanghaiTech University. Z. Wang was supported by the National Nature Science Foundation of China (Grant No. 11974395), the Strategic Priority Research Program of Chinese Academy of Sciences (Grant No. XDB33000000), and the Center for Materials Genome. We thank the staffs from BL15U1 at Shanghai Synchrotron Radiation Facility, for assistance during data collection. The calculations were carried out at the HPC Platform of ShanghaiTech University Library and Information Services, the SBP project, and the School of Physical Science and Technology.

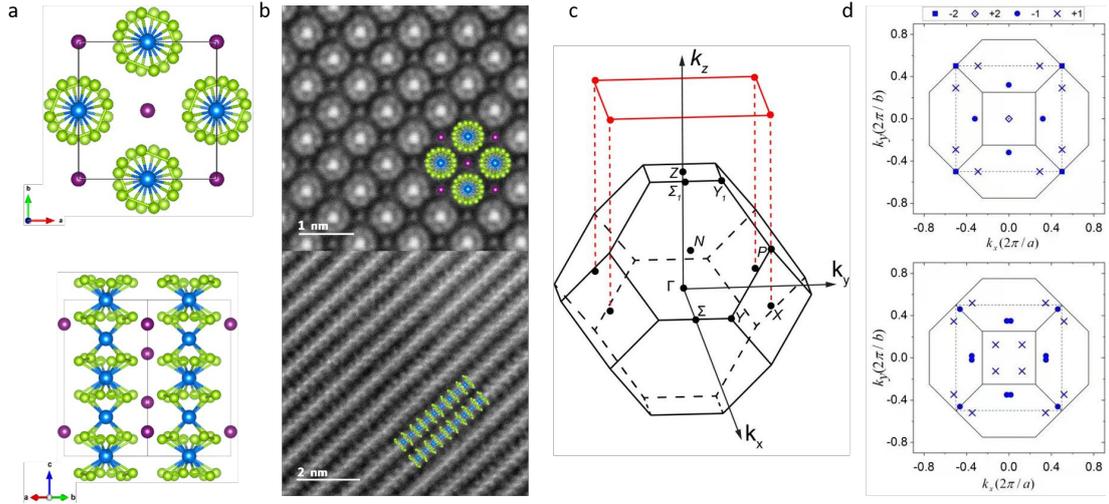

**Figure 1.** The conventional crystal structure of (NbSe$_4$)$_2$I, high-angle annular dark-field scanning transmission electron microscopy (HAADF-STEM) image, Brillouin zones (BZs), and the distribution of WPs. a) The conventional crystal structure is shown from both top (001) and tilted side perspectives, the blue, green, and purple symbols stand for Nb, Se, and I atoms, respectively. b) HAADF-STEM image of (NbSe$_4$)$_2$I. c) The bulk BZ of primitive cell and its projections onto the (001) surfaces. d) The distribution of WPs without (up) and with SOC (down) in the first bulk BZ, projected on the (001) surface BZ. Because of the bulk crystal symmetries, each symbol represents a pair of WPs with the same Chern number lying at the opposite momenta $\pm k_z$.

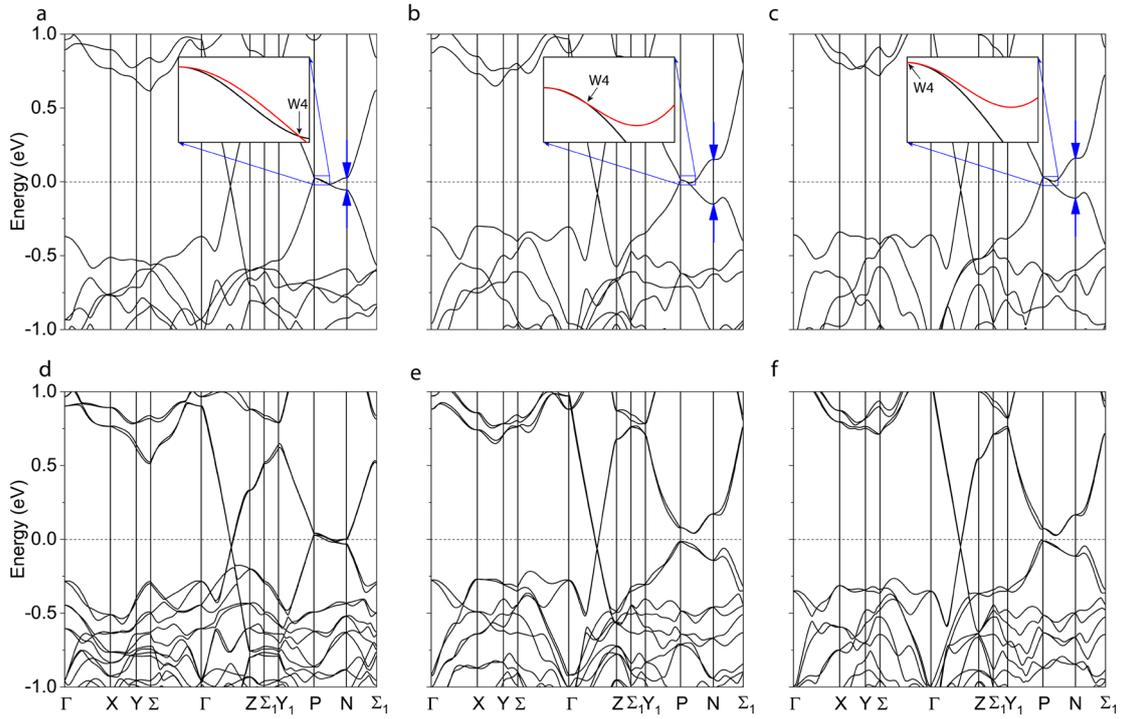

**Figure 2.** The calculated electronic band structure of (NbSe$_4$)$_2$I. a-c) band structure at ambient pressure, 5.0 GPa, and 9.7 GPa without SOC, respectively. d-f) band structure at the same pressure with SOC, respectively. A symmetry-protected charge-2 WP (W3) is pinned at *P*. The insets in a-c are the zoom in along the *P-N* direction. It is clear that the Weyl point W4 moves close to *P* as increasing pressure. With SOC included, the Weyl points move away from the *P-N* line, resulting a band gap in the *P-N* direction. The Fermi level is set to zero, and represent as the dashed line.

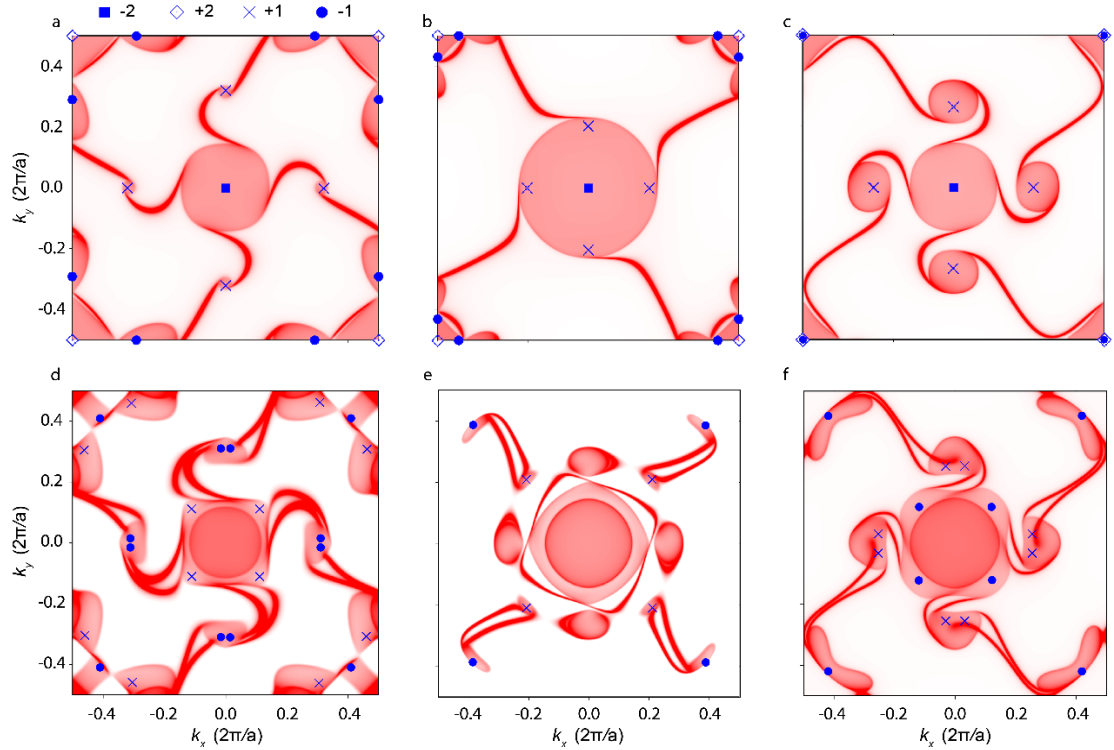

**Figure 3.** The calculated Fermi surface of (001) surface of conventional cell. a-c) the Fermi surface calculated without SOC at ambient pressure, 5.0 GPa, 9.7 GPa, respectively. d-f) the Fermi surface calculated with SOC at the same condition. Due to the crystal and time reversal symmetry, all the WPs projected on (001) direction are two separated WPs in the full BZ. It is clear that Fermi arcs connect the separated island which hosts the WPs. The number of arcs from the separated islands are identity to the Chern number in every island.

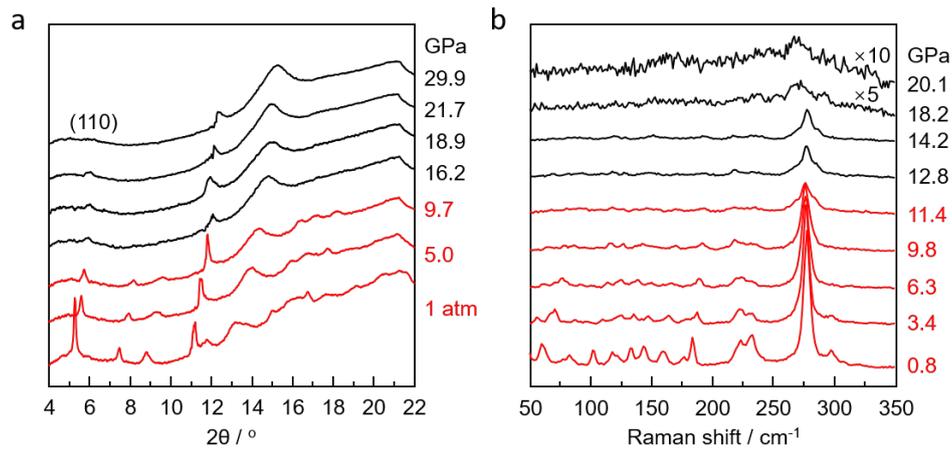

**Figure 4.** Pressure effect on structure of $(NbSe_4)_2I$. a) XRD patterns collected at various pressures for $(NbSe_4)_2I$ with an X-ray wavelength $\lambda = 0.6199$ Å. b) Raman spectrum of $(NbSe_4)_2I$ at various pressures.

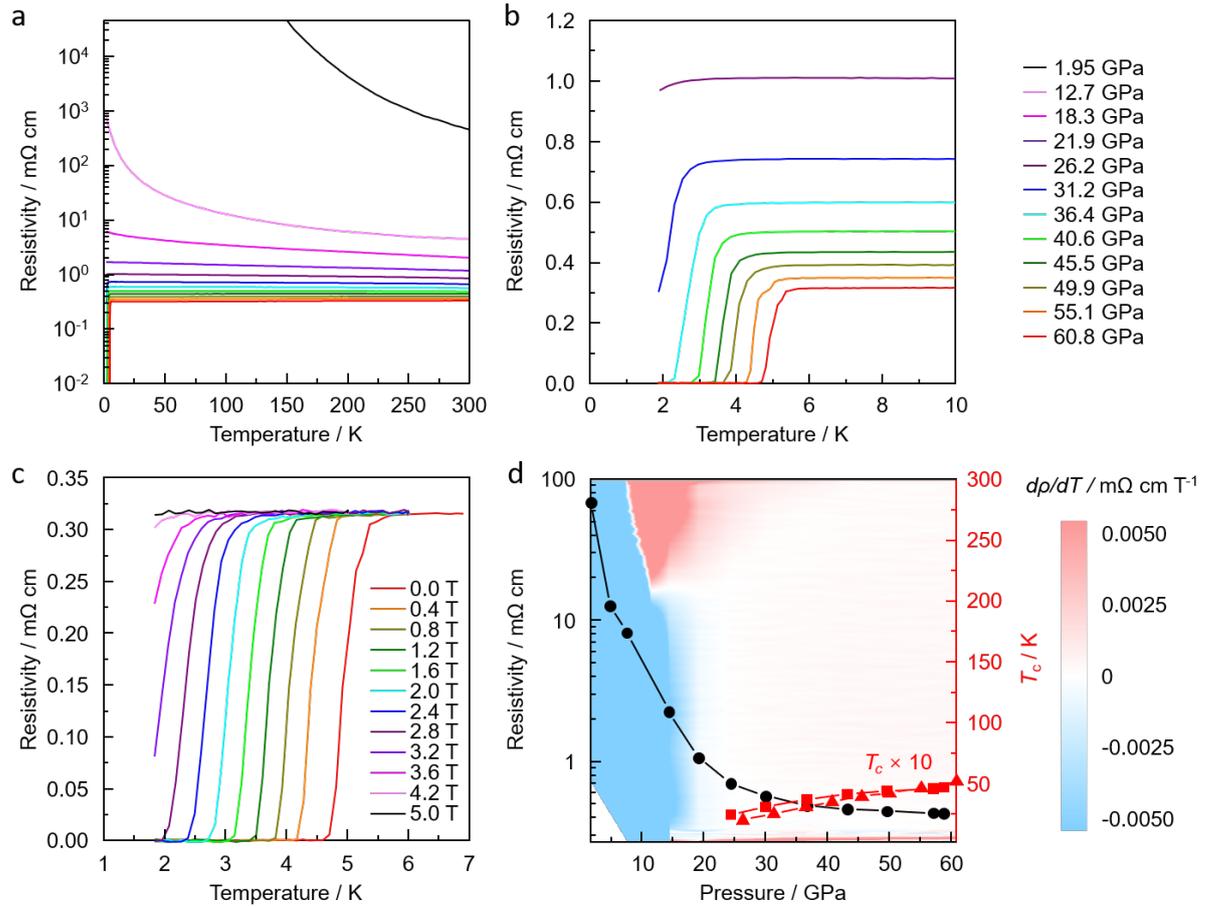

**Figure 5.** Emergence of superconductivity in the quasi-1D (NbSe$_4$)$_2$I. a) Electrical resistivity of (NbSe$_4$)$_2$I as a function of temperature at various pressures; b) Temperature-dependent resistivity of (NbSe$_4$)$_2$I in the vicinity of the superconducting transition; c) Temperature dependence of resistivity under different magnetic fields for (NbSe$_4$)$_2$I at 60.8 GPa; d) Superconducting $T_c$ (red symbols) and resistivity at 300 K (black symbols) of (NbSe$_4$)$_2$I as a function of pressure. The solid circles and the open circles represent data from two different runs of measurements.

**Table 1.** The calculated symmetry related Weyl points in (NbSe$_4$)$_2$I at ambient pressure, 5.0 GPa, and 9.7 GPa with and without SOC, respectively. With SOC, at ambient pressure (5.0 GPa, 9.7 GPa), the WPs are divided into four (two, three) symmetry related sets, while without SOC, at ambient pressure, 5.0 GPa, and 9.7 GPa, the Weyl points are divided into four symmetry related sets. Due to the crystal symmetries ($C_{4z}$, $C_{2x/y}$) and time reversal symmetry, all the general WPs are 16 symmetry equal points, in the $k_x=k_y$, $k_x=0$, and $k_y=0$ plane, the WPs are 8 symmetry equal points, along the $k_x=k_y=0$, or $k_x=k_y=0.5$ line, the WPs are 2 symmetry equal points. The positions are given in units of ($2\pi/a$, $2\pi/a$, $2\pi/c$) in Cartesian coordinates.

| | Pressure | | Coordinate ($2\pi/a$, $2\pi/a$, $2\pi/c$) | $E-E_F$ /meV | Chern number | Multiplicity |
|---|---|---|---|---|---|---|
| SOC | Ambient | WP1 | (0.111, 0.111, 0.413) | 8.608 | +1 | 8 |
| | | WP2 | (0.308, 0.462, 0.484) | 2.220 | +1 | 16 |
| | | WP3 | (0.410, 0.410, 0.481) | 18.692 | -1 | 8 |
| | | WP4 | (0.311, 0.015, 0.428) | 22.258 | -1 | 16 |
| | 5.0 GPa | WP1 | (0.209, 0.209, 0.384) | -34.749 | +1 | 8 |
| | | WP2 | (0.386, 0.386, 0.438) | -44.999 | -1 | 8 |
| | 9.7 GPa | WP1 | (0.255, 0.031, 0.380) | 25.188 | +1 | 16 |
| | | WP2 | (0.419, 0.419, 0.461) | 0.770 | -1 | 8 |
| | | WP3 | (0.120, 0.120, 0.382) | -9.114 | -1 | 8 |
| Without SOC | Ambient | WP1 | (0.000, 0.000, 0.405) | -30.483 | -2 | 2 |
| | | WP2 | (0.500, 0.500, 0.500) | 23.884 | +2 | 2 |
| | | WP3 | (0.000, 0.320, 0.429) | 4.668 | +1 | 8 |
| | | WP4 | 0.500, 0.291, 0.500) | -12.682 | -1 | 8 |
| | 5.0 GPa | WP1 | (0.000, 0.000, 0.402) | -79.429 | -2 | 2 |
| | | WP2 | (0.500, 0.500, 0.500) | 12.749 | +2 | 2 |
| | | WP3 | (0.000, 0.203, 0.391) | -16.383 | +1 | 8 |
| | | WP4 | (0.500, 0.430, 0.500) | 4.747 | -1 | 8 |
| | 9.7 GPa | WP1 | (0.000, 0.000, 0.376) | -55.672 | -2 | 2 |
| | | WP2 | (0.500, 0.500, 0.500) | 28.029 | +2 | 2 |
| | | WP3 | (0.000, 0.265, 0.375) | 32.191 | +1 | 8 |
| | | WP4 | (0.500, 0.499, 0.500) | 28.029 | +1 | 8 |

# Supporting information

**Experimental Section**

***Sample synthesis and characterization in ambient condition***: The $(NbSe_4)_2I$ single crystals used in this work were grown using a chemical vapor transport method with Nb, Se and I as starting materials[35]. The phase and quality examinations were performed on a Bruker D8 single crystal X-ray diffractometer (SXRD) with Mo $K_{\alpha 1}$ ($\lambda$ = 0.71073 Å) at 298 K. The TEM and STEM samples were prepared by a focused ion beam (JEOL JIB-4700F) and further processed by a low energy ion slicer (Technoorg Linda Gentle Mill Model IV8). Thickness of the samples are down to about less than 100 nm. The high-resolution STEM images and atomic EDS mappings were obtained on a 300 kV electron microscope equipped with a cold field emission gun (FEG) and double spherical aberration correctors (JEOL GrandARM300F). Angle-resolved photoemission spectroscopy (ARPES) measurement was carried out using a lab-based laser source with a photon energy of 6.994 eV.

***High pressure resistivity measurements***: High-pressure resistivity measurements were performed in a nonmagnetic diamond anvil cell (DAC). A cubic BN/epoxy mixture layer was inserted between BeCu gaskets and electrical leads. Four Pt foil were arranged with a van der Pauw four-probe method to contact the sample in the chamber for resistivity measurements. Pressure was determined by the ruby luminescence method[37]. A magnet-cryostat (Dynacool, Quantum Design, $T_{min}$ = 1.8 K) was used to take the cryogenic setup and magnetic field measurements.

***High pressure Raman spectroscopy measurements***: *In situ* high-pressure Raman spectroscopy investigation on $(NbSe_4)_2I$ was carried out on Raman spectrometer (Renishaw inVia, U.K.) with a laser excitation wavelength of 532 nm and low-wavenumber filter. Symmetric DAC with anvil culet sizes of 300 μm as well as silicon oil was used as pressure transmitting medium (PTM).

***High pressure XRD measurements***: *In situ* high-pressure XRD measurements were performed at the beamline 15U at Shanghai Synchrotron Radiation Facility (X-ray wavelength $\lambda$ = 0.6199 Å). Symmetric DAC with anvil culet sizes of 250 μm as well as

Re gaskets were used. Silicon oil was used as PTM and pressure was determined by the ruby luminescence method[2]. The two-dimensional diffraction images were analyzed using the FIT2D software[38]. Rietveld refinements on crystal structures under high pressure were performed by General Structure Analysis System (GSAS) and graphical user interface EXPGUI package[39-40].

***Theoretical calculation***: The first principle calculations were performed based on density functional theory as implemented in Vienna *Ab initio* Simulation Package (VASP)[41] with the projector augmented wave potential[42-43]. The exchange-correlation potential was formulated by generalized gradient approximation with Perdew-Burke-Ernzerhof (PBE) functional[44]. A $\Gamma$-center 8×8×8 *k* points grid was used for the first Brillouin zone sampling. Crystal structures determined from XRD experiment (ambient pressure, 5.0 GPa, and 9.7 GPa) were used in all the calculations. The tight-binding Hamiltonian was constructed by projecting atomic Wannier functions[45] on the Nb 4*d*, Se 4*p*, and I 5*p* orbitals to reproduce the calculated band structures. The surface states were obtained by calculating the surface Green's functions[46-47] of a semi-infinite tight-binding model constructed from the above Wannier functions.

**Table S1.** Crystallographic Data for $(NbSe_4)_2I$

| | |
|---|---|
| (1) Physical, crystallographic and analytical data | |
| Formula | $(NbSe_4)_2I$ |
| Formula weight (g/mol) | 912.67 |
| Temperature (K) | 150.0 |
| Crystal system | Tetragonal |
| Space group | I422 |
| a (Å) | 9.4896(4) |
| b (Å) | 9.4896(4) |
| c (Å) | 12.7724(5) |
| α (°) | 90 |
| β (°) | 90 |
| γ (°) | 90 |
| $V$ (Å$^3$) | 1150.19(11) |
| Z | 4 |
| Density (calc., g / cm$^3$) | 5.271 |
| Crystal description | Needle |
| Crystal size (mm$^3$) | 1 × 0.1 × 0.1 |
| (2) Data collection | |
| Radiation | MoKα (λ = 0.71073) |
| Reflections collected | 11532 |
| Index ranges | -12 ≤ h ≤ 13 |
| | -13 ≤ k ≤ 12 |
| | -18 ≤ l ≤ 18 |
| 2$\theta$ range (°) | 5.348 - 63.934 |
| (3) Data reduction and Refinement | |
| $\mu$ (mm$^{-1}$) | 29.260 |
| F (000) | 1575.0 |
| Independent reflections | 917 [$R_{int}$ = 0.0774, $R_{sigma}$ = 0.0437] |
| Data/restraints/parameters | 917/13/36 |
| Goodness-of-fit on F$^2$ | 1.096 |
| Largest diff. peak/hole / e Å$^{-3}$ | 1.03/-0.97 |
| Flack parameter | 0.06(2) |
| Final R indexes [I>=2σ (I)] | $R_1$ = 0.0290, w$R_2$ = 0.0579 |
| Final R indexes [all data] | $R_1$ = 0.0520, w$R_2$ = 0.0624 |

**Table S2.** Fractional Atomic Coordinates (×10⁴) and Equivalent Isotropic Displacement Parameters (Å²×10³) for (NbSe$_4$)$_2$I.

| Atom | x | y | z | U(eq) |
|---|---|---|---|---|
| Se1 | -0.7180.4(9) | 427.5(9) | -6306.0(7) | 27.7(2) |
| Nb1 | -0.5000 | 0 | -5000 | 26.1(4) |
| Se2 | -0.6227.2(8) | -1881.8(9) | -6189.7(7) | 27.7(2) |
| Nb2 | -0.5000 | 0 | -7500 | 26.4(4) |
| I1 | -0.5000 | -5000 | -6559.0(15) | 27.7(5) |
| I2 | -0.5000 | -5000 | -4710(20) | 16(5) |
| I3 | -0.5000 | -5000 | -5000 | 27(6) |

**Table S3.** Anisotropic Displacement Parameters (Å²×10³) for (NbSe$_4$)$_2$I.

| Atom | U$_{11}$ | U$_{22}$ | U$_{33}$ | U$_{23}$ | U$_{13}$ | U$_{12}$ |
|---|---|---|---|---|---|---|
| Se1 | 25.3(4) | 27.5(4) | 30.5(4) | -0.8(4) | -0.3(4) | 1.1(4) |
| Nb1 | 22.6(9) | 25.1(9) | 30.7(8) | 0 | 0 | 0 |
| Se2 | 26.5(4) | 25.9(4) | 30.6(4) | -0.6(4) | 1.2(4) | -0.8(3) |
| Nb2 | 24.2(6) | 24.2(6) | 30.8(8) | 0 | 0 | -0.3(6) |
| I1 | 25.5(6) | 25.5(6) | 32.1(10) | 0 | 0 | 0 |
| I2 | 14(6) | 14(6) | 20(11) | 0 | 0 | 0 |
| I3 | 26(5) | 26(5) | 27(12) | 0 | 0 | 0 |

Noted: The Anisotropic displacement factor exponent takes the form:
-2π²[h²a*²U$_{11}$+2hka*b*U$_{12}$+…].

### $k \cdot p$ Model at $P$ point

To illustrate the feature of the chirality W2 WP at $P$ point, we consider a $k \cdot p$ Hamiltonian without and with SOC. Under the little-group of $P$ point (i.e. $C_{2z}$, $C_{2y}$, and $TC_{4z}$), the chirality C=2 WPs can be captured by following two band $k \cdot p$ model[34]

$$H_{nsoc}(k) = A(k_x^2 + k_y^2)\tau_0 + B(k_x^2 - k_y^2)\tau_z + Ck_z^2\tau_0 + d_1 k_z\tau_x + d_2 k_z\tau_y + e_1 k_x k_y \tau_x$$
$$+ e_2 k_x k_y \tau_y + fk_z(k_x^2 + k_y^2)\tau_x$$

where the coefficients $(A, B, C, d_1, d_2, e_1, e_2, f)$ are real and $\tau_i (i = 1, 2, 3)$ are Pauli matrices. By fitting the PBE band structure of (NbSe$_4$)$_2$I, these parameters are given in Table I. Once including SOC effect, the charge-2 Weyl point becomes four monopole Weyl points away from $P$ point, which is captured by our four-band $k \cdot p$ model. The effective model is derived as,

$$H(k) = \sigma_0 \otimes H_{nsoc}(k) + \lambda H_{soc}(k)$$

where

$$H_{soc}(k) = c_1\sigma_z \otimes \tau_x + c_2 k_z \sigma_z \otimes \tau_0 + c_3(k_x\sigma_x + k_y\sigma_y) \otimes \tau_0 + c_4(k_x\sigma_x - k_y\sigma_y) \otimes \tau_z$$
$$+ c_5(k_y\sigma_x + k_x\sigma_y) \otimes \tau_x$$

$\sigma_{x,y,z}$ are spin Pauli matrices. By fitting the PBE+SOC band structures, the SOC parameters are obtained in Table 2. When $\lambda = 0.05$, the position of four monopole Weyl points (offset from $P$ point) are $(\pm 0.0483, \pm 0.0483, 0.0075)$ and $(\mp 0.0483, \pm 0.0483, -0.0075)$ in units of $1/\text{Å}$.

**Table S4.** Fitting parameters for the nosoc $k \cdot p$ model.

| $A(\text{eVÅ}^2)$ | $B(\text{eVÅ}^2)$ | $C(\text{eVÅ}^2)$ | $d_1(\text{eVÅ}^2)$ | $d_2(\text{eVÅ}^2)$ | $e_1(\text{eVÅ}^2)$ | $e_2(\text{eVÅ}^2)$ | $f(\text{eVÅ}^2)$ |
|---|---|---|---|---|---|---|---|
| -2.98 | 0.5 | 0.0 | -1.88 | 4.42 | 5.91 | -7.77 | 37.07 |

| $c_1(\text{eV})$ | $c_2(\text{eVÅ})$ | $c_3(\text{eVÅ})$ | $c_4(\text{eVÅ})$ | $c_5(\text{eVÅ})$ |
|---|---|---|---|---|
| 0.020 | 0.042 | 0.002 | 0.12 | 0.0 |

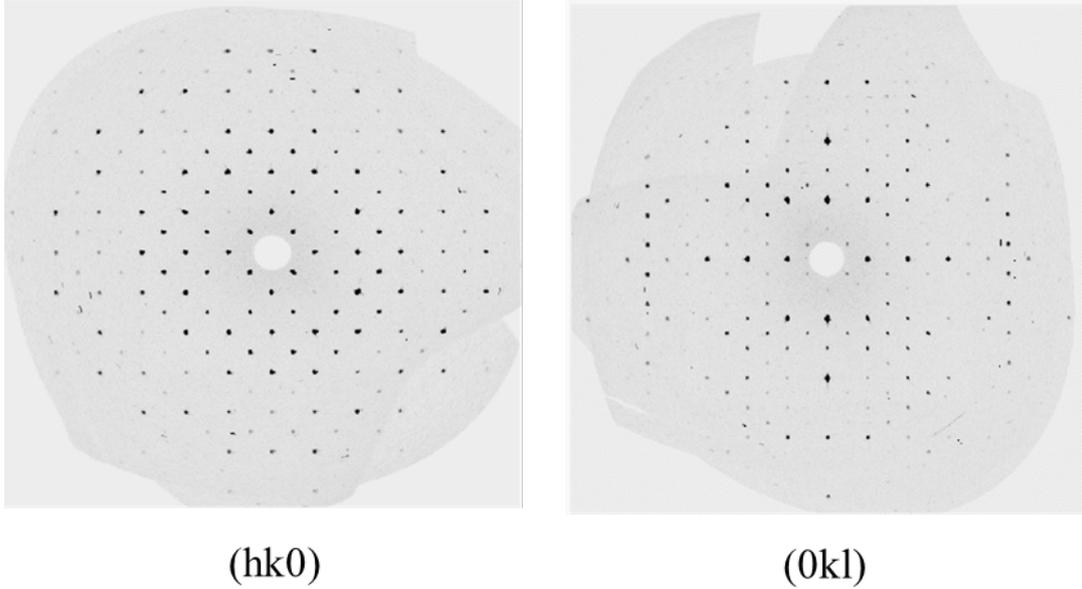

(hk0)　　　　　　　　　　　　(0kl)

**Figure S1.** Diffraction patterns of $(\text{NbSe}_4)_2\text{I}$ in the reciprocal space along (h k 0) and (0 k l) directions. The diffraction pattern can be well indexed on the basis of a tetragonal structure in the space group *I*422 (No. 97). The details of crystallographic data are shown in Table S1-3.

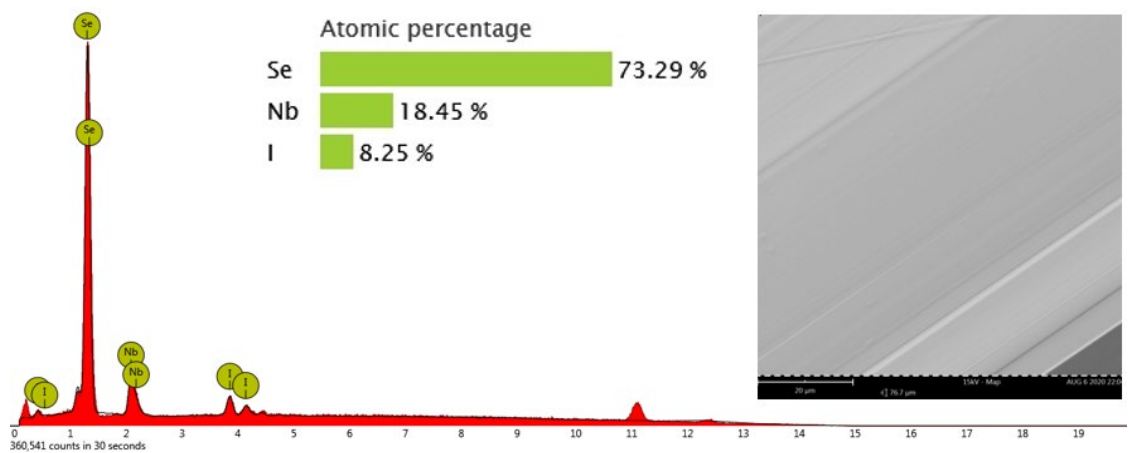

**Figure S2.** The stoichiometry of (NbSe$_4$)$_2$I crystal measured by the EDS spectrum. The picture shows the crystal used for EDS measurements. The average compositions derived from a typical EDS measured at several points on the crystal, revealing good stoichiometry with the atomic ratio of Nb : Se : I = 18.45 : 73.29 : 8.25.

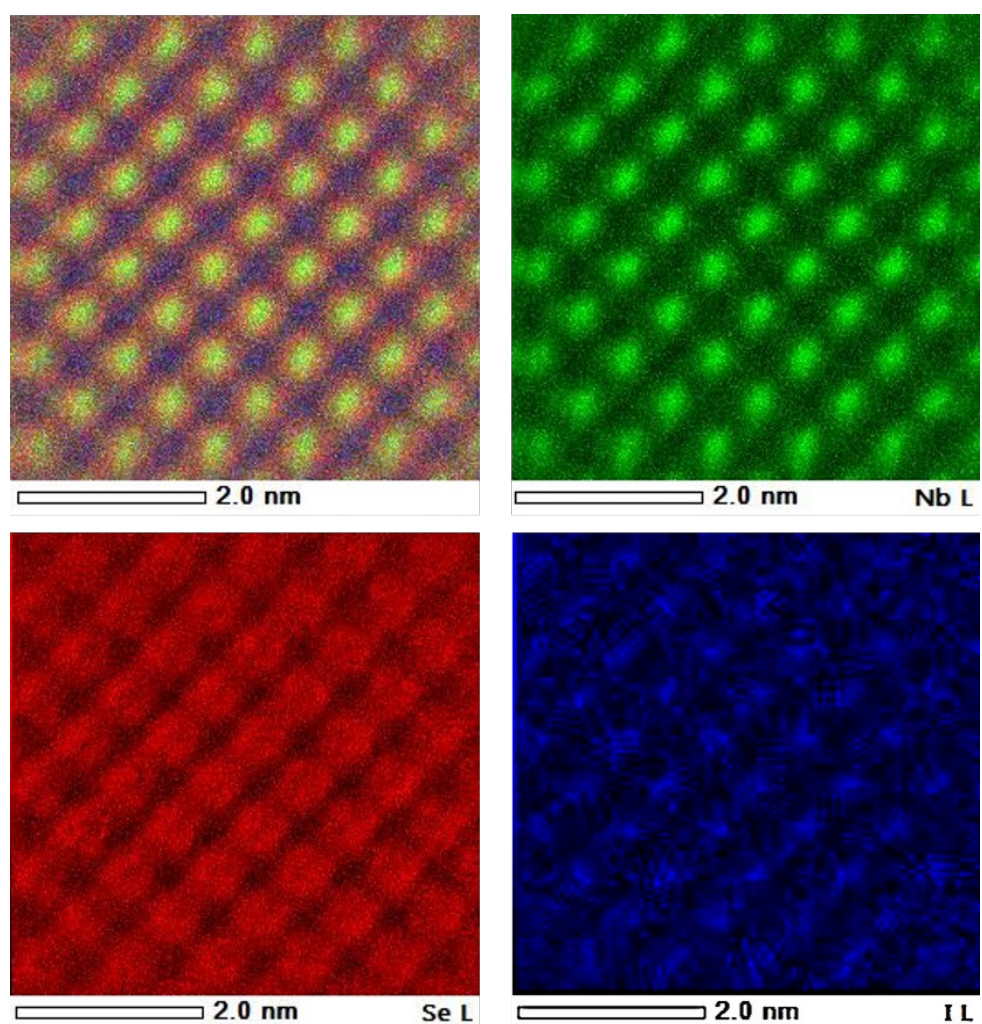

**Figure S3.** Atomic scale elemental mapping of (NbSe$_4$)$_2$I.

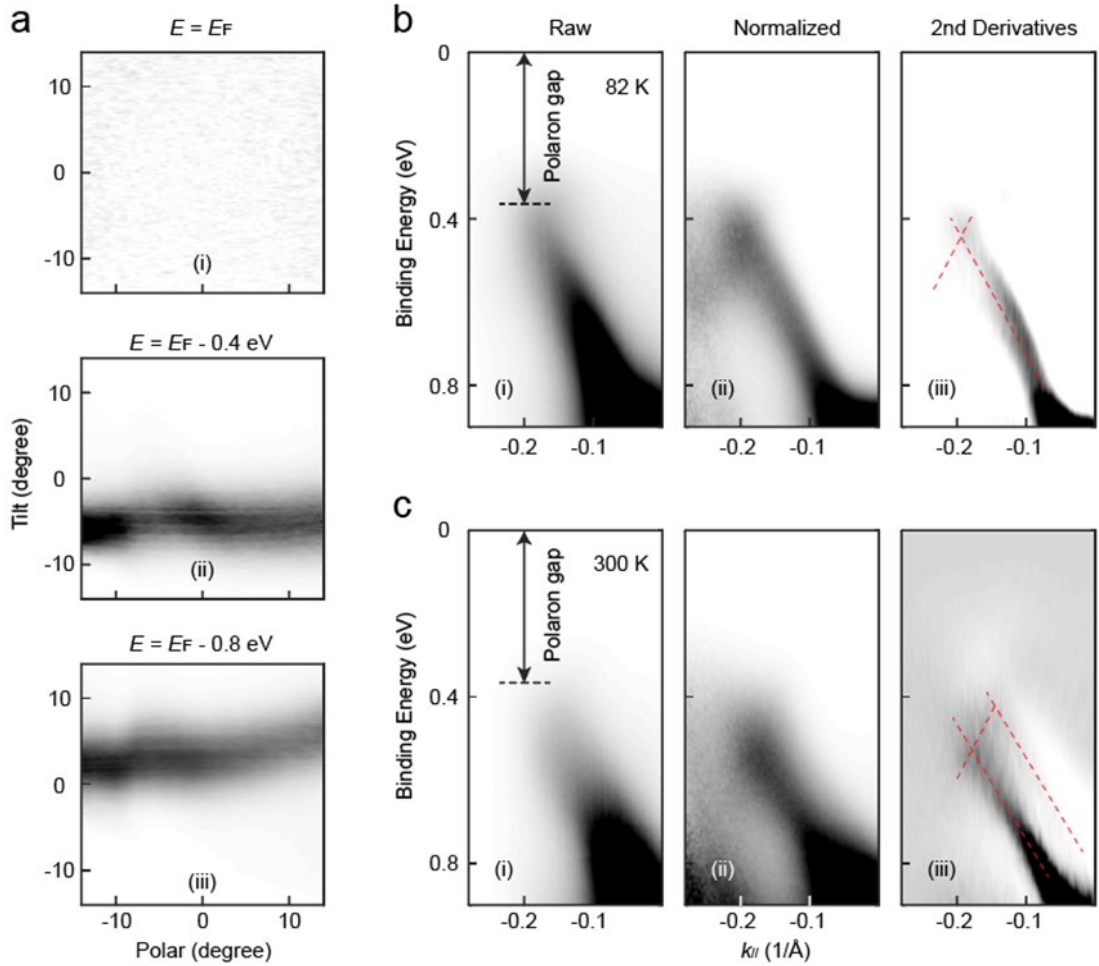

Figure S4. ARPES measurements on (NbSe$_4$)$_2$I. (a), Equal energy mapping at three binding energies. (i) $E = E_F$; (ii) $E = E_F - 0.4\ eV$; (iii) $E = E_F - 0.8\ eV$. (b), Band dispersion measured at 82 K. (i) Raw ARPES spectrum. (ii) Normalized ARPES spectrum along the momentum distribution curve (MDC) direction. (iii) Second derivatives along the energy distribution curve (EDC) direction. (c), Same as (b) but measured at 300 K.

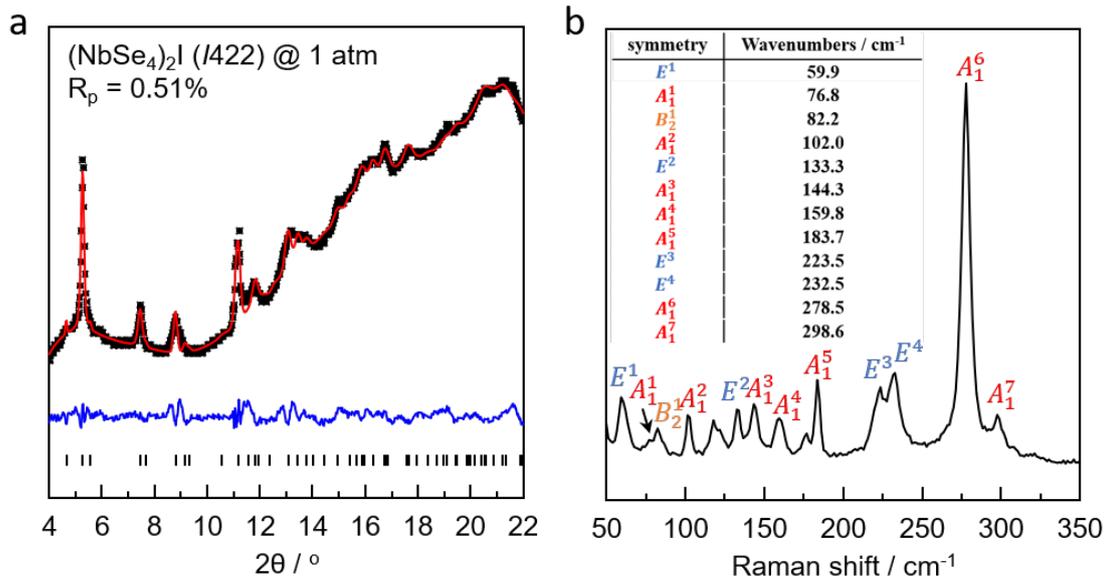

**Figure S5.** a) Typical Rietveld refinement of (NbSe$_4$)$_2$I at 1 atm. The experimental and calculated patterns are indicated by black stars and red lines, respectively. The solid lines shown at the bottom of the figures are the residual intensities. The vertical bars indicate peak positions of the Bragg reflections for (NbSe$_4$)$_2$I in *I*422 space groups. b) Raman spectrum of (NbSe$_4$)$_2$I at 0.7 GPa with peak assigned to $A_1$, $B_2$ and $E$ vibration modes.

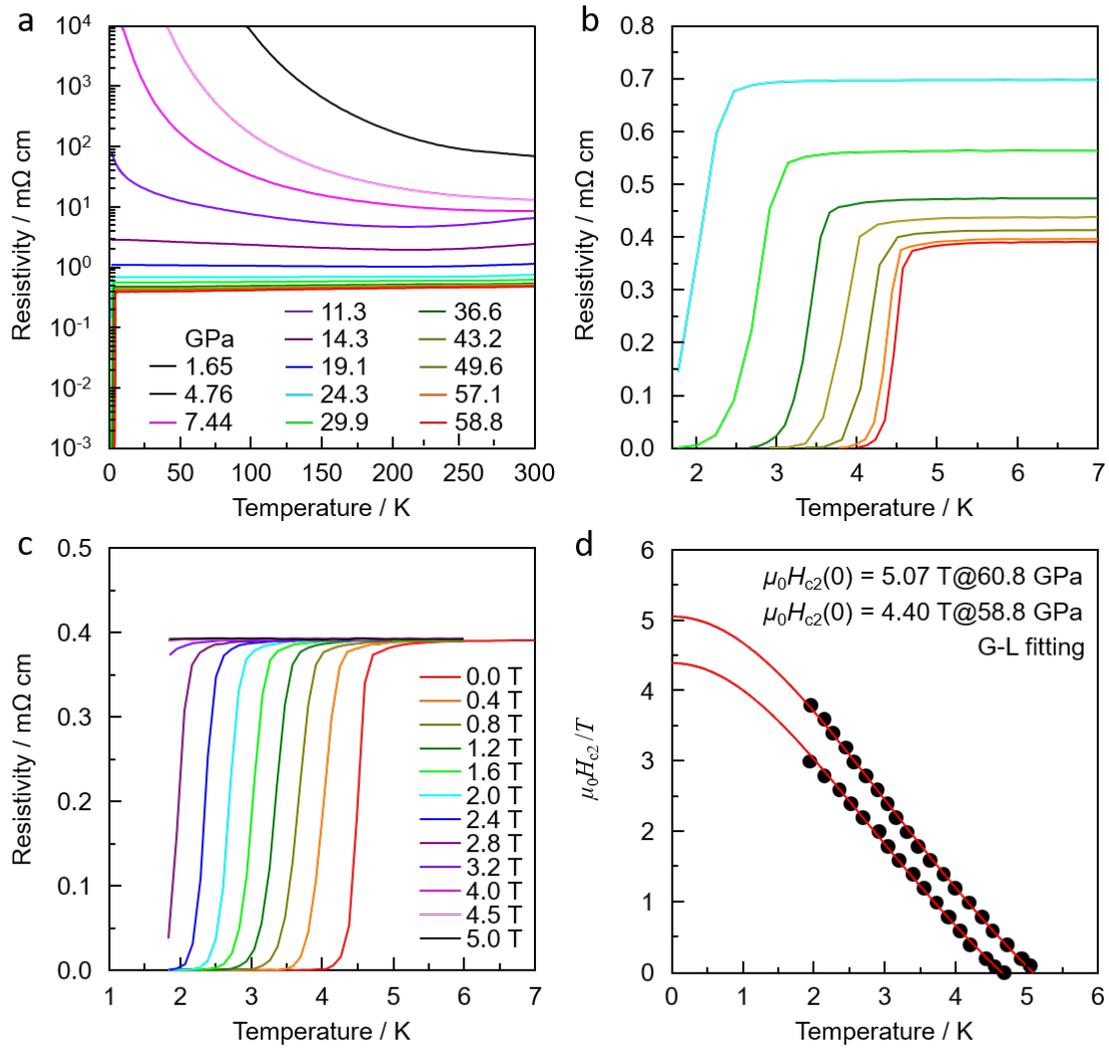

**Figure S6.** a) Electrical resistivity $\rho(T)$ of (NbSe$_4$)$_2$I as a function of temperature for pressures up to 58.8 GPa in run II. b) Enlarged $\rho(T)$ curves in the vicinity of the superconducting transition. Zero resistivity is obtained for pressure over 29.9 GPa, indicating the emergence of superconductivity. c) Temperature dependence of resistivity under different magnetic fields for (NbSe$_4$)$_2$I at 58.8 GPa in run II. d) Temperature dependence of upper critical field for (NbSe$_4$)$_2$I at 58.8 GPa and 60.8 GPa (run I), respectively. Here, $T_c$ is determined as the 90% drop of the normal state resistivity. The solid lines represent the fits based on the Ginzburg–Landau (GL) formula.

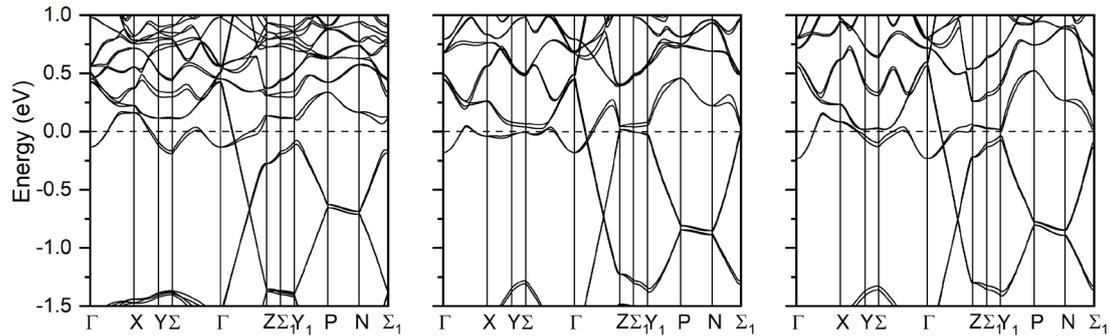

**Figure S7.** The calculated electronic band structure of (NbSe$_4$)$_2$I with I atoms removed using the lattice constant at ambient pressure, 5.0 GPa, and 9.7 GPa, respectively. The SOC is considered. Compared with Fig. 2 (d-f), the difference is the Fermi level shift to higher energy, this means that the I atoms accept electrons from the NbSe$_4$-chain.

## References

1. Grüner, G., The dynamics of charge-density waves. *Rev. Mod. Phys.* **1988,** *60* (4), 1129-1181.
2. Voit, J., One dimensional Fermi liquids. *Rep. Prog. Phys.* **1995,** *58*, 977.
3. Giamarchi, T., Quantum physics in one dimension. *Oxford University Press, Oxford, England* **2004**.
4. Monceau, P., Electronic crystals: an experimental overview. *Adv. Phys.* **2012,** *61*, 325.
5. Haldane, F. D. M., Luttinger liquid theory of one dimensional quantum fluids I properties of the Luttinger model and their extension to the gneral 1D interacting spinless Fermi gas. *J. Phys. C: Solid State Phys.* **1981,** *14*, 2585.
6. P.Gressier; A.Meerschaut; L.Guemas; J.Rouxel, Characterization of the new series of quasi one-dimensional compounds(MX$_4$)$_n$Y (M =Nb, Ta; X =S, Se; Y =Br, I). *J. Solid State Chem.* **1984,** *51*, 141-151.
7. Lorenzo, J. E.; Currat, R.; Monceau, P.; Hennion, B.; Berger, H.; Levy, F., A neutron scattering study of the quasi-one-dimensional conductor (TaSe$_4$)$_2$I. *J. Phys.: Condens. Matter.* **1998,** *10*, 5039.
8. Gressier, P.; Guemas, L.; Meerschaut, A., Preparation and structure of ditantalum iodide octaselenide Ta$_2$ISe$_8$. *Acta Crystallogr., Sect. B: Struct. Crystallogr. Cryst. Chem.* **1982,** *38*, 2877.
9. Gooth, J.; Bradlyn, B.; Honnali, S.; Schindler, C.; Kumar, N.; Noky, J.; Qi, Y.; Shekhar, C.; Sun, Y.; Wang, Z.; Bernevig, B. A.; Felser, C., Axionic Charge-Density Wave in the Weyl Semimetal (TaSe$_4$)$_2$I. *Nature* **2019,** *575* (7782), 315-319.
10. Shi, W.; Wieder, B. J.; Meyerheim, H. L.; Sun, Y.; Zhang, Y.; Li, Y.; Shen, L.; Qi, Y.; Yang, L.; Jena, J.; Werner, P.; Koepernik, K.; Parkin, S.; Chen, Y.; Felser, C.; Bernevig, B. A.; Wang, Z., A charge-density-wave topological semimetal. *Nat. Phys.* **2021,** *17*, 381-387.
11. Cava, R. J.; Littlewood, P.; Fleming, R. M.; Dunn, R. G.; Rietman, E. A., Low-frequency dielectric response of the charge-density wave in (TaSe$_4$)$_2$I. *Phys Rev B Condens Matter* **1986,** *33* (4), 2439-2443.
12. Smaalen, S. V.; Lam, E. J.; Ludecke, J., Structure of the charge density wave in (TaSe$_4$)$_2$I. *J. Phys.: Condens. Matter.* **2001,** *13*, 9923.
13. Li, X.-P.; Deng, K.; Fu, B.; Li, Y.; Ma, D.-S.; Han, J.; Zhou, J.; Zhou, S.; Yao, Y., Type-III Weyl semimetals: (TaSe$_4$)$_2$I. *Phys. Rev. B* **2021,** *103* (8), L081402.
14. Zhang, Y.; Lin, L.-F.; Moreo, A.; Dong, S.; Dagotto, E., First-principles study of the low-

temperature charge density wave phase in the quasi-one-dimensional Weyl chiral compound (TaSe$_4$)$_2$I. *Phys. Rev. B* **2020,** *101* (17), 174106.

15. Yi, H.; Huang, Z.; Shi, W.; Min, L.; Wu, R.; Polley, C. M.; Zhang, R.; Zhao, Y.-F.; Zhou, L.-J.; Adell, J.; Gui, X.; Xie, W.; Chan, M. H. W.; Mao, Z.; Wang, Z.; Wu, W.; Chang, C.-Z., Surface charge induced Dirac band splitting in a charge density wave material (TaSe$_4$)$_2$I. *Phys. Rev. Res.* **2021,** *3* (1), 013271.

16. Philipp, A.; Mayr, W., Dynamical charge density wave response in (NbSe$_4$)$_2$I. *Solid State Commun.* **1987,** *62*, 521-524.

17. Sugai, S.; Sato, M.; Kurihara, S., Interphonon interactions at the charge-density-wave phase transitions in (TaSe4)2I and (NbSe4)2I. *Phys. Rev. B Condens Matter* **1985,** *32* (10), 6809-6818.

18. Qi, Y.; Shi, W.; Naumov, P. G.; Kumar, N.; Schnelle, W.; Barkalov, O.; Shekhar, C.; Borrmann, H.; Felser, C.; Yan, B.; Medvedev, S. A., Pressure dirven superconductivity in the transition metal pentatelluride HfTe$_5$. *Phys. Rev. B* **2016,** *94*, 054517.

19. Petrovic, A. P.; Ansermet, D.; Chernyshov, D.; Hoesch, M.; Salloum, D.; Gougeon, P.; Potel, M.; Boeri, L.; Panagopoulos, C., A disorder-enhanced quasi-one-dimensional superconductor. *Nat. Commun.* **2016,** *7*, 12262.

20. Li, X.; Chen, D.; Jin, M.; Ma, D.; Ge, Y.; Sun, J.; Guo, W.; Sun, H.; Han, J.; Xiao, W.; Duan, J.; Wang, Q.; Liu, C. C.; Zou, R.; Cheng, J.; Jin, C.; Zhou, J.; Goodenough, J. B.; Zhu, J.; Yao, Y., Pressure-induced phase transitions and superconductivity in a quasi-1-dimensional topological crystalline insulator alpha-Bi4Br4. *Proc Natl Acad Sci U S A* **2019,** *116* (36), 17696-17700.

21. Qi, Y.; Shi, W.; Werner, P.; Naumov, P. G.; Schnelle, W.; Wang, L.; Rana, K. G.; Parkin, S.; Medvedev, S. A.; Yan, B.; Felser, C., Pressure induced superconductivity and topological quantum phase transitions in a quasi one dimensional topological insulator Bi$_4$I$_4$. *npj Quantum Mater.* **2018,** *3*, 4.

22. Zhang, J.; Jia, Y.; Wang, X.; Li, Z.; Duan, L.; Li, W.; Zhao, J.; Cao, L.; Dai, G.; Deng, Z.; Zhang, S.; Feng, S.; Yu, R.; Liu, Q.; Hu, J.; Zhu, J.; Jin, C., A new quasi-one-dimensional compound Ba$_3$TiTe$_5$ and superconductivity induced by pressure. *NPG Asia Mater.* **2019,** *11* (1), 60.

23. Jérome, D., Organic conductors: from charge density wave TTF-TCNQ to superconducting (TMTSF)$_2$PF$_6$. *Chem. Rev.* **2004,** *104*, 5565-5591.

24. Köhler, B.; Rose, E.; Dumm, M.; Untereiner, G.; Dressel, M., Comprehensive transport study of anisotropy and ordering phenomena in quasi-one-dimensional(TMTTF)$_2$X salts (X = PF$_6$, AsF$_6$, SbF$_6$, BF$_4$, ClO$_4$, ReO$_4$). *Phys. Rev. B* **2011,** *84* (3), 035124.

25. Rose, E.; Loose, C.; Kortus, J.; Pashkin, A.; Kuntscher, C. A.; Ebbinghaus, S. G.; Hanfland, M.; Lissner, F.; Schleid, T.; Dressel, M., Pressure-dependent structural and electronic properties of quasi-one-dimensional (TMTTF)$_2$PF$_6$. *J. Phys.: Condens. Matter.* **2013,** *25* (1), 014006.

26. Moser, J.; Gabay, M.; Auban-Senzier, P.; Jérome, D.; Bechgaard, K.; Fabre, J. M., Transverse transport in (TM)$_2$X organic conductors: possible evidence for a Luttinger liquid. *Eur. Phys. J. B* **1998,** *1*, 39-46.

27. Itoi, M.; Kano, M.; Kurita, N.; Hedo, M.; Uwatoko, Y.; Nakamura, T., Pressure-induced superconductivity in the quasi-one-dimensional organic conductor (TMTTF)$_2$AsF$_6$. *J. Phys. Soc. Jpn.* **2007,** *76* (5), 053703.

28. Vuletić, T.; Auban-Senzier, P.; Pasquier, C.; Tomić, S.; Jérome, D.; Héritier, M.; Bechgaard, K., Coexistence of superconductivity and spin density wave orderings in the organic superconductor (TMTSF)$_2$PF$_6$. *Eur. Phys. J. B* **2002,** *25* (3), 319-331.

29. Narayanan, A.; Kiswandhi, A.; Graf, D.; Brooks, J.; Chaikin, P., Coexistence of spin density waves and superconductivity in (TMTSF)$_2$PF$_6$. *Phys. Rev. Lett.* **2014,** *112* (14), 146402.

30. Pisoni, A.; Gaál, R.; Zeugner, A.; Falkowski, V.; Isaeva, A.; Huppertz, H.; Autès, G.; Yazyev, O. V.; Forró, L., Pressure effect and superconductivity in the β−Bi$_4$I$_4$ topological insulator. *Phys. Rev. B* **2017,** *95* (23), 235147.

31. Forró, L.; Mutka, H.; Bouffard, S.; Morillo, J.; Jánossy, A., Ohmic and nonlinear transport of (TaSe$_4$)$_2$I under pressure. *Lect. Notes Phys.* **1985,** *217*, 360-365.

32. Mu, Q.; Nenno, D.; Qi, Y.; Fan, F.; Pei, C.; ElGhazali, M.; Gooth, J.; Felser, C.; Narang, P.; Medvedev, S., Suppression of axionic charge density wave and onset of superconductivity in the chiral Weyl semimetal Ta$_2$Se$_8$I. **2020**, arXiv:2010.07345.

33. An, C.; Zhou, Y.; Chen, C.; Fei, F.; Song, F.; Park, C.; Zhou, J.; Rubahn, H. G.; Moshchalkov, V. V.; Chen, X.; Zhang, G.; Yang, Z., Long-Range Ordered Amorphous Atomic Chains as Building Blocks of a Superconducting Quasi-One-Dimensional Crystal. *Adv. Mater.* **2020,** *32*, 2002352.

34. Liu, Q.-B.; Qian, Y.; Fu, H.-H.; Wang, Z., Symmetry-enforced Weyl phonons. *npj Computational Materials* **2020,** *6* (1), 95.

35. Fujishita, H.; Sato, M.; Sato, S.; Hoshino, S., X ray diffraction study of the quasi one dimensional conductors MSe4 2I M Ta and Nb. *J. Phys. C: Solid State Phys.* **1985,** *18*, 1105.

36. Kusmartseva, A. F.; Sipos, B.; Berger, H.; Forro, L.; Tutis, E., Pressure induced superconductivity in pristine 1T-TiSe$_2$. *Phys. Rev. Lett.* **2009,** *103* (23), 236401.

37. Mao, H. K.; Xu, J.; Bell, P. M., Calibration of the ruby pressure gauge to 800 kbar under quasi-hydrostatic conditions. *J. Geophys. Res.* **1986,** *91* (B5), 4673-4676.

38. Hammersley, A. P.; Svensson, S. O.; Hanfland, M.; Fitch, A. N.; Hausermann, D., Two-dimensional detector software: From real detector to idealised image or two-theta scan. *High Press. Res.* **1996,** *14* (4-6), 235-248.

39. Larson, A. C.; Dreele, R. B. V., General structure analysis system (GSAS). *Los Alamos National Laboratory Report LAUR* **2004**, 86-748.

40. Toby, B., EXPGUI, a graphical user interface for GSAS. *J. Appl. Crystallogr.* **2001,** *34* (2), 210-213.

41. Kresse, G.; Furthmüller, J., Efficient iterative schemes for ab initio total-energy calculations using a plane-wave basis set. *Phys. Rev. B* **1996,** *54* (16), 11169-11186.

42. Blochl, P. E., Projector augmented-wave method. *Phys Rev B Condens Matter* **1994,** *50* (24), 17953-17979.

43. Kresse, G.; Joubert, D., From ultrasoft pseudopotentials to the projector augmented wave method. *Phys. Rev. B* **1999,** *59*, 1758.

44. Perdew, J. P.; Burke, K.; Ernzerhof, M., Generalized gradient approximation made simple. *Phys. Rev. Lett.* **1996,** *77*, 3865.

45. Mostofi, A. A.; Yates, J. R.; Lee, Y.-S.; Souza, I.; Vanderbilt, D.; Marzari, N., Wannier90: A tool for obtaining maximally-localised Wannier functions. *Comput. Phys. Commun.* **2008,** *178*, 685-699.

46. Sancho, M. P. L.; Sancho, J. M. L.; Rubio, J., Quick iterative scheme for the calculation of transfer matrices application to Mo (100). *J. Phys. F* **1984,** *14*, 1205-1215.

47. Sancho, M. P. L.; Sancho, J. M. L.; Rubio, J., Highly convergent schemes for the calculation of bulk and surface Green functions. *J. Phys. F* **1985,** *15*, 851-858.